\documentclass[onecolumn]{IEEEtran}
\usepackage[utf8]{inputenc}
\usepackage{amsmath,bm}
\usepackage{amsthm}
\usepackage{amssymb}
\usepackage{mathtools}
\allowdisplaybreaks
\usepackage{graphicx}
\usepackage{xcolor}
\usepackage[english]{babel}
\usepackage{tabularx}
\usepackage{multirow}
\usepackage{cite}

\usepackage{tikz}\usetikzlibrary{shapes.multipart}
\usetikzlibrary{positioning}
\tikzset{font={\fontsize{9pt}{12}\selectfont}}
\tikzset{>=latex}
\usetikzlibrary{calc}
\usetikzlibrary{arrows}
\usetikzlibrary{external}
\usetikzlibrary{patterns}
\usetikzlibrary{fit}
\usepackage[draft]{changes}

\usetikzlibrary{automata,arrows}

\DeclareMathOperator{\ir}{Irr}

\DeclareMathOperator{\mds}{MDS}

\newtheorem{theorem}{Theorem$\!$}
\newtheorem{example}[theorem]{Example$\!$}
\newtheorem{lemma}[theorem]{Lemma$\!$}
\newtheorem{corollary}[theorem]{Corollary$\!$}

\newtheorem{construction}[theorem]{Construction$\!$}

\usepackage{graphicx}

\usepackage{xcolor}

\newcommand{\bcomment}[1]{{\leavevmode\color{blue}#1}}

\newcommand{\mA}{\mathsf{A}}
\newcommand{\mC}{\mathsf{C}}
\newcommand{\mG}{\mathsf{G}}
\newcommand{\mT}{\mathsf{T}}

\newcommand{\cB}{\mathcal{B}}
\newcommand{\cC}{\mathcal{C}}
\newcommand{\cD}{\mathcal{D}}

\newcommand{\cL}{\mathcal{L}}

\newcommand{\bbF}{\mathbb{F}}

\newcommand{\mybold}[1]{\bm{#1}}
\newcommand{\ba}{{\mybold{a}}}

\newcommand{\bc}{{\mybold{c}}}

\newcommand{\br}{{\mybold{r}}}
\newcommand{\bs}{{\mybold{s}}}
\newcommand{\bt}{{\mybold{t}}}
\newcommand{\bu}{{\mybold{u}}}
\newcommand{\bv}{{\mybold{v}}}
\newcommand{\bw}{{\mybold{w}}}
\newcommand{\bx}{{\mybold{x}}}
\newcommand{\by}{{\mybold{y}}}
\newcommand{\bz}{{\mybold{z}}}

\newcommand{\balpha}	{\mybold{\alpha}}
\newcommand{\bbeta}		{\mybold{\beta}}
\newcommand{\bgamma}	{\mybold{\gamma}}

\newcommand{\bsigma}	{\mybold{\sigma}}

\IEEEoverridecommandlockouts
\title{Error-correcting Codes for Short Tandem Duplication and Substitution Errors}

\author{
  \IEEEauthorblockN{Yuanyuan Tang 
                     and Farzad Farnoud
                     }
                     \\
  \IEEEauthorblockA{
                    Electrical \& Computer Engineering,
                    University of Virginia,
                    \texttt{\{yt5tz,farzad\}@virginia.edu}} \\
  \thanks{This work was supported in part by NSF grants under grant nos.~1816409 and~1755773
  . This paper was presented in part at the 2020 IEEE Symposium of Information Theory (ISIT) in 2020 \cite{tang2020error}.}
}
\begin{document}

\maketitle

\begin{abstract}
Due to its high data density and longevity, DNA  is considered a promising  medium for satisfying ever-increasing data storage needs. However, the diversity of errors that occur in DNA sequences makes efficient error-correction a challenging task. This paper aims to address simultaneously correcting two types of errors, namely, short tandem duplication and substitution errors. We focus on tandem repeats of length at most 3 and design codes for correcting an arbitrary number of duplication errors and one substitution error. Because a substituted symbol can be duplicated many times (as part of substrings of various lengths), a single substitution can affect an unbounded substring of the retrieved word. However, we show that with appropriate preprocessing, the effect may be limited to a substring of finite length, thus making efficient error-correction possible. We construct a code for correcting the aforementioned errors and provide lower bounds for its rate. Compared to optimal codes correcting only duplication errors, numerical results show that the asymptotic cost of protecting against an additional substitution is only $0.003$ bits/symbol when the alphabet has size 4, an important case corresponding to data storage in DNA.
\end{abstract}

\section{Introduction}
Recent advances in DNA synthesis and sequencing technologies~\cite{yazdi2015dna} have made DNA a promising candidate for rising data storage needs. Compared to traditional storage media, DNA storage has several advantages, including higher data density, longevity, and ease of generating copies~\cite{yazdi2015dna}. 
However, DNA is subject to a diverse set of errors that may occur during the various stages of data storage and retrieval, including substitutions, duplications, insertions, and deletions. This poses a challenge to the design of error-correcting codes and has led to many recent works studying the subject, including~\cite{yazdi2015dna,jain2017duplication, kovavcevic2018asymptotically, yehezkeally2019reconstruction, tang2020single,lenz2019coding, cai2019optimal, elishco2019bounds,lenz2020coding, gabrys2020mass,jain2020coding,kiah2020coding,yehezkeally2020uncertainty, nguyen2020constrained, sima2020robust}. 
The current paper focuses on correcting short duplication  and substitution errors.


A (tandem) duplication error generates a copy of a substring of the DNA sequence and inserts it after the original substring~\cite{jain2017duplication}. For example, from $\mA\mC\mG\mT$ we may obtain $\mA\mC\mG\underline{\mC\mG}\mT$. The length of the duplication is the length of the substring being copied, which is 2 in the preceding example. In the literature, both fixed-length duplication \cite{jain2017duplication,yehezkeally2019reconstruction,kovavcevic2018asymptotically, tang2020single} and bounded-length duplication, where the duplication length is bounded from above \cite{jain2017duplication, jain2017capacity, kovavcevic2019codes, chee2019deciding} have been studied.  {For duplications whose length is at most 3, the case most relevant to this paper, Jain et al.~\cite{jain2017duplication} proposed error-correcting codes that were shown to have an asymptotically optimal rate by Kova{\v{c}}evi{\'c}~\cite{kovavcevic2019codes}.}

In a substitution event, a symbol in the sequence is changed to another alphabet symbol. 
Substitution errors may be restricted to the inserted copies, reflecting the noisiness of the copying mechanism during the duplication process~\cite{pumpernik2008replication,farnoud2018} or be unrestricted. For fixed-length duplication, these settings have been studied in~\cite{tang2020single,tang2019error}.



We focus on correcting errors that may arise from channels with many duplication errors of length at most 3, which we refer to as \emph{short duplications}, and one unrestricted substitution error. Considering a single substitution error reveals important insights into the interactions between substitution and duplication errors and will be of use for studying the general case of $t$ substitution errors. As a simple example of this channel, the input $\mA\mC\mG$ may become $\mA\mC\mT\mC\mT\mA\mC\mT\mA\mC\mT\mC\mG$, where the occurrences of the symbol $\mT$ result from copies of the substitution $\mC\to\mT$. Given that an arbitrary number of duplications are possible, an unbounded segments of the output word may be affected by the errors and the incorrect, substituted symbol may appear many times. However, relying on the fact that short tandem duplications lead to regular languages, We show that with an appropriate construction and preprocessing of the output of the channel, the deleterious effects of the errors may be localized.
We leverage constrained coding and maximum distance separable codes to design codes for correcting the resulting errors, establish a lower bound on the code rate, and provide an asymptotic analysis that shows that the code has rate at least $\log (q-2)$, where $q$ is the size of the alphabet and the $\log$ is in base 2. We note that the rate of the code correcting only short duplications is upper bounded by $\log (q-1)$. When $q=4$, the case corresponding to DNA storage, we provide a computational bound for the code rate, showing that asymptotically its rate is only $0.003$ bits/symbol smaller than the code that corrects short duplications but no substitutions.

The paper is organized as follows. In Section \ref{sec:Not_Pre}, we provide the notation and relevant background. Section \ref{Sec:channel_model} analyzes the errors patterns that result from passing through  duplication and substitution channels. After that, the code construction as well as the code size are presented in Section \ref{sec:ECC}. Finally, Section~\ref{sec:conclusion} presents our concluding remarks. 

\section{Notation and Preliminaries}\label{sec:Not_Pre}
Let $\Sigma_q=\{0,1,\dotsc,q-1\}$ denote a finite alphabet of size $q$.
To avoid trivial cases, we assume $q\ge3$, which in particular includes the case of $q=4$, relevant to DNA data storage.
The set of all strings of finite length  over $\Sigma_q$ is denoted by $\Sigma_q^{*}$, while $\Sigma_q^{n}$ represents the strings of length $n$. In particular, $\Sigma_q^*$ contains the empty string $\Lambda$. Let $[n]$ denote the set $\{1,\dotsc,n\}$.

Strings over $\Sigma_q$ are denoted by bold symbols, such as $\bx$ and $\by_j$, or by capital letters. The elements of strings are shown with plain typeface, e.g., $\bx=x_1x_2\dotsm x_n$ and $\by_j=y_{j1}y_{j2}\dotsm y_{jm}$, where $x_i,y_{ji}\in\Sigma_q$. Given two strings $\bx, \by\in \Sigma_q^{*}$, $\bx\by$ denotes their concatenation and  $\bx^{m}$ denotes the concatenation of $m$ copies of $\bx$. We use $|\bx|$ to denote the length of a word $\bx\in \Sigma_q^{*}$. For four words $\bx,\bu,\bv,\bw \in \Sigma_{q}^{*}$, if $\bx$ can be expressed as $\bx=\bu\bv\bw$, then $\bv$ is a \emph{substring} of $\bx$.

Given a word $\bx\in \Sigma_{q}^{*}$, a \emph{tandem duplication} (TD) of length $k$ copies a substring of length $k$ and inserts it after the original. This is referred to as a $k$-TD. For example, a 2-TD may generate $abcbcde$ from $abcde$. Here, $bcbc$ is called a \emph{(tandem) repeat} of length 2. Our focus in this paper is on TDs of length bounded by $k$, denoted $\leq\!\!k$-TD, for $k=3$. For example, given $\bx=1201210$ we may obtain via $\le\!\!3$-TDs
\begin{equation}
\begin{split}
    \label{eq:exam_upperk_TD}
        \bx=&1201210\to1201\underline{201}210\to\\
        &120120\underline{20}1210\to 1201202\underline{2}01210=\bx',
\end{split}
\end{equation}
where the underlined substrings are the inserted copies. We say that $\bx'$ is a \emph{descendant} of $\bx$, i.e., a sequence resulting from $\bx$ through a sequence of duplications.

Let {$\ir_{\leq k}(n) \subseteq \Sigma_{q}^{n}$} denote the set of \emph{irreducible strings} (more precisely, $\le\!\!k$-irreducible strings) of length $n$, i.e., strings without repeats of length at most $k$. We use $\ir_{\leq k}(*)$ denotes $\le\!\! k$-irreducible strings of arbitrary lengths. Furthermore, let $D_{\leq k}^*(\bx)$ denote the \emph{descendant cone} of $\bx$, containing all the descendants of $\bx$ after an arbitrary number of $\leq\!\! k$-TDs. 
Given a string $\bx$, let
\begin{equation*}
    R_{\leq k}(\bx) = \{\br\in\ir_{\le k}(*)|\bx\in D_{\leq k}^{*}(\br)\}
\end{equation*}
denote the set of \emph{duplication root}s of $\bx$, i.e., repeat-free sequences of which $\bx$ is a descendant. For a set $S$ of strings, $R_{\leq k}(S)$ is the set of strings each of which is a root of at least one string in $S$. If $R_{\leq k}(\cdot)$ is a singleton, we may view it as a string rather than a set.  A root can be obtained from $\bx$ by repeatedly replacing all repeats of the form $\ba\ba$ with $\ba$, where $|\ba|\le k$ (each such operation is called a \emph{deduplication}). For $\le\!\!3$-TDs, the duplication root is unique~\cite{jain2017duplication}
. If $\bx'$ is a descendant of $\bx$, we have $R_{\le3}(\bx)= R_{\le3}(\bx')$. 
For $k=3$, we may drop the $\le3$ subscript from the notation and write $D^*(\cdot),R(\cdot),\ir(\cdot)$.

We also consider substitution errors, although our attention is limited to at most one error of this kind.
Continuing the example given in~\eqref{eq:exam_upperk_TD}, a substitution occurring in the descendant $\bx'$ of $\bx$ may result in $\bx''$:
 \begin{equation}
    \begin{split}\label{eq:Example_upperk_TD_sub}
         \bx'=1201202201210 &\rightarrow \bx''=1201202\underline{1}01210.
    \end{split}
\end{equation}
We denote by $D_{\leq k}^{t,p}(\bx)$ the set of strings that can be obtained from $\bx$ through $t$  TDs of length at most $k$ and $p$ substitutions, \emph{in any order}. We note that  substitutions are \emph{unrestricted} in the sense that they may occur in any position in the string, unlike the \emph{noisy duplication} setting, where they are restricted to the inserted copies \cite{tang2020single, tang2019error}. 
Replacing $t$ with $*$ denotes any number of $\le\!\!k$-TDs and replacing $p$ with $\le p$ denotes at most $p$ substitutions. We again drop $\le k$ from the notation when $k=3$. In the example given in~\eqref{eq:exam_upperk_TD} and~\eqref{eq:Example_upperk_TD_sub}, we have $\bx''\in D^{*,1}(\bx)$, 
denoting that $\bx''$ is a descendant generated from $\bx$ after an arbitrary number of $\le\!\!3$-TDs and a substitution error. 

\section{Channels with many $\le\!\!3$-TDs\\ and one substitution error}\label{Sec:channel_model}
In this section, we study channels that alter the input string by applying an arbitrary number of duplication errors and at most one substitution error, where the substitution may occur at any time in the sequence of errors. We will first study the conditions a code must satisfy to be able to correct such errors. Then, we will investigate the effect of such channels on the duplication root of sequences, which is an important aspect of designing our error-correcting codes.

A code $C$ is able to correct an arbitrary number of $\le\!\!3$-TDs and a substitution if and only if for any two distinct codewords $\bc_{1}, \bc_{2}\in C$, we have
\[
    D^{*,\leq 1}(\bc_1)\cap D^{*,\leq 1}(\bc_2)= \varnothing.
\]
To satisfy this condition, it is sufficient to have
\begin{equation}\label{eq:R}
        R(D^{*,\leq 1}(\bc_1))\cap R(D^{*,\leq 1}(\bc_2))= \varnothing.
\end{equation}
Condition~\eqref{eq:R} implies that for distinct codewords $\bc_1$ and $\bc_2$, $R(\bc_1)\ne R(\bc_2)$. This latter condition is in fact sufficient for correcting only $\le\!\!3$-TDs since this type of error does not alter the duplication root. For correcting only $\le\!\!3$-TDs, defining the code as the set of irreducible strings of a given length leads to asymptotically optimal codes~\cite{jain2017duplication,kovavcevic2019codes}. The decoding process is simply finding the root of the received word.

We take a similar approach to correct many $\le\!\!3$-TDs and a substitution. 
More specifically, the proposed code $C$ is a subset of $\le\!\!3$-irreducible strings, i.e., $R(\bc)=\bc$ for $\bc\in C$. To recover $\bc$ from the received word $\by$, we find $R(\by)$ and from that recover $R(\bc)=\bc$, as will be discussed.

We start by studying the effect of $\le\!\!3$-TDs and one substitution on the root of a string. Specifically, for strings $\bx$ and $\bx''\in D^{*,\leq 1}(\bx)$, it is of interest to determine how $R(\bx'')$ differs from $R(\bx)$. We either have $\bx''\in D^*(\bx)$, i.e., $\bx''$ suffers only duplications, or $\bx''\in D^{*,1}(\bx)$. In the former case $R(\bx'')=R(\bx)$. Hence, below we consider only $\bx''\in D^{*,1}(\bx)$. Note that duplications that occur after the substitution do not affect the root and so in our analysis we may assume that the substitution is the last error. We start by a lemma that considers a simple case. 


\begin{figure*}
    \begin{center}
        \begin{tikzpicture}[thin, scale=0.8, transform shape, vnd/.style={shape=circle,draw,inner sep=0,outer sep=0, minimum width=1cm, scale=1},
        invertibleedge/.style={->, line width=.3mm}]
            \node (Start) [vnd] {$Start$};
            \node (S1) [ right = of Start, vnd] {$S_1$};
            \node (S2) [ right = of S1, vnd] {$S_2$};
            \node[state,accepting] (S3) [ below right = of S2, vnd] {$S_3$};
            \node (S4) [ above right = of S3, vnd] {$S_4$};
            \node (T2) [ above = of S2, vnd] {$T_2$};
            \node (T4) [ above = of S4, vnd] {$T_4$};
            \node (T3) [ below= of S3, vnd] {$T_3$};

            \draw [->] (Start) -- (S1) node [midway, above, rotate=0] {$0$};
             \draw [->] (S1) -- (S2) node [midway, above, rotate=0] {$1$};
             \draw [->] (S2) -- (S3) node [midway, above, rotate=0] {$2$};
             \draw [->] (S3) -- (S4) node [midway, above, rotate=0] {$0$};
            \draw [->] (S4) -- (S2) node [midway, above, rotate=0] {$1$};

           \draw [->] (S1) edge [out=70,in=110,looseness=6] node[above] {0} (S1);
          \draw [->] (S2) edge [out=-70,in=-110,looseness=6] node[below] {1} (S2);
          \draw [->] (T2) edge [out=70,in=110,looseness=6] node[above] {0} (T2);
          \draw [->] (T4) edge [out=70,in=110,looseness=6] node[above] {2} (T4);
           \draw [->] (T3) edge [out=-70,in=-110,looseness=6] node[below] {1} (T3);

         \draw [->] (S3) edge [out=70,in=110,looseness=6] node[above] {2} (S3);
        \draw [->] (S4) edge [out=-70,in=-20,looseness=6] node[below] {0} (S4);

            \draw [->] (S4) to [out=50,in=-50]  node [midway, left, rotate=0]{$2$} (T4) ;
            \draw [->] (T4) to [out=230,in=130]  node [midway, right, rotate=0]{$0$} (S4);
            \draw [->] (S2) to [out=50,in=-50]  node [midway, left, rotate=0]{$0$} (T2) ;
            \draw [->] (T2) to [out=230,in=130]  node [midway, right, rotate=0]{$1$} (S2);
             \draw [->] (T3) to [out=50,in=-50]  node [midway, left, rotate=0]{$2$} (S3) ;
            \draw [->] (S3) to [out=230,in=130]  node [midway, right, rotate=0]{$1$} (T3);

            \end{tikzpicture}
    \caption{Finite automaton for the regular language $D^*(012)$ based on~\cite{jain2017capacity}. } \label{fig:FA_IS}
    \end{center}
\end{figure*}
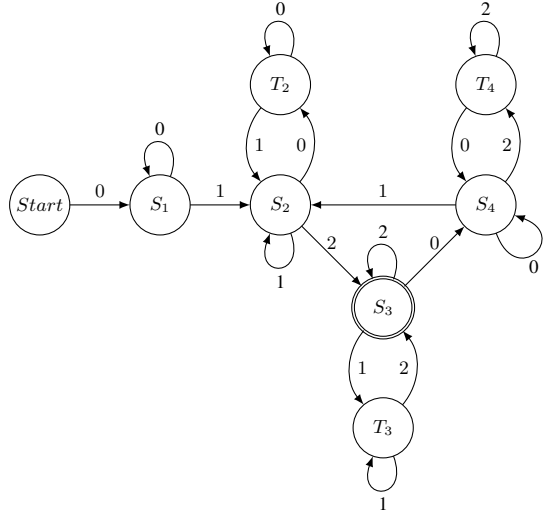


\begin{figure*}
    \begin{center}
        \begin{tikzpicture}[thin, scale=0.8, transform shape, vnd/.style={shape=circle,draw,inner sep=0,outer sep=0, minimum width=1cm, scale=1},
        invertibleedge/.style={->, line width=.3mm}]
            \node (Start) [vnd] {$Start$};
            \node (S1) [ right = of Start, vnd] {$S_1$};
            \node (S2) [ right = of S1, vnd] {$S_2$};
            \node (S3) [ below right = of S2, vnd] {$S_3$};
            \node (S4) [ above right = of S3, vnd] {$S_4$};
            \node (T2) [ above = of S2, vnd] {$T_2$};
            \node (T4) [ above = of S4, vnd] {$T_4$};
            \node (T3) [ below= of S3, vnd] {$T_3$};

            \node (S5) [ right = of S4, xshift=1cm, vnd] {$S_5$};
            \node (S6) [ below right = of S5, vnd] {$S_6$};
            \node (S7) [ above right = of S6, vnd] {$S_7$};
            \node (T5) [ above = of S5, vnd] {$T_5$};
            \node (T7) [ above = of S7, vnd] {$T_7$};
            \node (T6) [ below= of S6, vnd] {$T_6$};

            \node (S8) [ right = of S7, xshift=1cm,vnd] {$S_8$};
            \node[state,accepting] (S9) [ below right = of S8, vnd] {$S_9$};
            \node (S10) [ above right = of S9, vnd] {$S_{10}$};
            \node (T8) [ above = of S8, vnd] {$T_8$};
            \node (T10) [ above = of S10, vnd] {$T_{10}$};
            \node (T9) [ below= of S9, vnd] {$T_9$};

            \draw [->] (Start) -- (S1) node [midway, above, rotate=0] {$0$};
             \draw [->] (S1) -- (S2) node [midway, above, rotate=0] {$1$};
             \draw [->] (S2) -- (S3) node [midway, above, rotate=0] {$2$};
             \draw [->] (S3) -- (S4) node [midway, above, rotate=0] {$0$};
            \draw [->] (S4) -- (S2) node [midway, above, rotate=0] {$1$};

           \draw [->] (S1) edge [out=70,in=110,looseness=6] node[above] {0} (S1);
          \draw [->] (S2) edge [out=-70,in=-110,looseness=6] node[below] {1} (S2);
          \draw [->] (T2) edge [out=70,in=110,looseness=6] node[above] {0} (T2);
          \draw [->] (T4) edge [out=70,in=110,looseness=6] node[above] {2} (T4);
           \draw [->] (T3) edge [out=-70,in=-110,looseness=6] node[below] {1} (T3);

         \draw [->] (S3) edge [out=70,in=110,looseness=6] node[above] {2} (S3);
        \draw [->] (S4) edge [out=-70,in=-20,looseness=6] node[below] {0} (S4);

            \draw [->] (S4) to [out=50,in=-50]  node [midway, left, rotate=0]{$2$} (T4) ;
            \draw [->] (T4) to [out=230,in=130]  node [midway, right, rotate=0]{$0$} (S4);
            \draw [->] (S2) to [out=50,in=-50]  node [midway, left, rotate=0]{$0$} (T2) ;
            \draw [->] (T2) to [out=230,in=130]  node [midway, right, rotate=0]{$1$} (S2);
             \draw [->] (T3) to [out=50,in=-50]  node [midway, left, rotate=0]{$2$} (S3) ;
            \draw [->] (S3) to [out=230,in=130]  node [midway, right, rotate=0]{$1$} (T3);

             \draw [->] (S3) -- (S6) node [midway, above, rotate=0] {$3$};
             \draw [->] (S6) -- (S7) node [midway, above, rotate=0] {$1$};
             \draw [->] (S7) -- (S5) node [midway, above, rotate=0] {$2$};
            \draw [->] (S5) -- (S6) node [midway, above, rotate=0] {$3$};

            \draw [->] (T5) edge [out=70,in=110,looseness=6] node[above] {1} (T5);
          \draw [->] (T7) edge [out=70,in=110,looseness=6] node[above] {3} (T7);
           \draw [->] (T6) edge [out=-70,in=-110,looseness=6] node[below] {2} (T6);

        \draw [->] (S6) edge [out=70,in=110,looseness=6] node[above] {3} (S6);
        \draw [->] (S7) edge [out=-70,in=-20,looseness=6] node[below] {1} (S7);
        \draw [->] (S5) edge [out=200,in=250,looseness=6] node[below] {2} (S5);

           \draw [->] (S7) to [out=50,in=-50]  node [midway, left, rotate=0]{$3$} (T7) ;
            \draw [->] (T7) to [out=230,in=130]  node [midway, right, rotate=0]{$1$} (S7);
            \draw [->] (S5) to [out=50,in=-50]  node [midway, left, rotate=0]{$1$} (T5) ;
            \draw [->] (T5) to [out=230,in=130]  node [midway, right, rotate=0]{$2$} (S5);
             \draw [->] (T6) to [out=50,in=-50]  node [midway, left, rotate=0]{$3$} (S6) ;
            \draw [->] (S6) to [out=230,in=130]  node [midway, right, rotate=0]{$2$} (T6);

             \draw [->] (S6) -- (S9) node [midway, above, rotate=0] {$4$};
             \draw [->] (S9) -- (S10) node [midway, above, rotate=0] {$2$};
             \draw [->] (S10) -- (S8) node [midway, above, rotate=0] {$3$};
            \draw [->] (S8) -- (S9) node [midway, above, rotate=0] {$4$};

           \draw [->] (T8) edge [out=70,in=110,looseness=6] node[above] {2} (T8);
          \draw [->] (T10) edge [out=70,in=110,looseness=6] node[above] {4} (T10);
           \draw [->] (T9) edge [out=-70,in=-110,looseness=6] node[below] {3} (T9);

          \draw [->] (S9) edge [out=70,in=110,looseness=6] node[above] {4} (S9);
        \draw [->] (S10) edge [out=-70,in=-20,looseness=6] node[below] {2} (S10);
        \draw [->] (S8) edge [out=200,in=250,looseness=6] node[below] {3} (S8);

           \draw [->] (S10) to [out=50,in=-50]  node [midway, left, rotate=0]{$4$} (T10) ;
            \draw [->] (T10) to [out=230,in=130]  node [midway, right, rotate=0]{$2$} (S10);
            \draw [->] (S8) to [out=50,in=-50]  node [midway, left, rotate=0]{$2$} (T8) ;
            \draw [->] (T8) to [out=230,in=130]  node [midway, right, rotate=0]{$3$} (S8);
             \draw [->] (T9) to [out=50,in=-50]  node [midway, left, rotate=0]{$4$} (S9) ;
            \draw [->] (S9) to [out=230,in=130]  node [midway, right, rotate=0]{$3$} (T9);

            \end{tikzpicture}
    \caption{Finite automaton for the regular language $D^*(01234)$ based on~\cite{jain2017capacity}. } \label{fig:fsm5}
    \end{center}
\end{figure*}
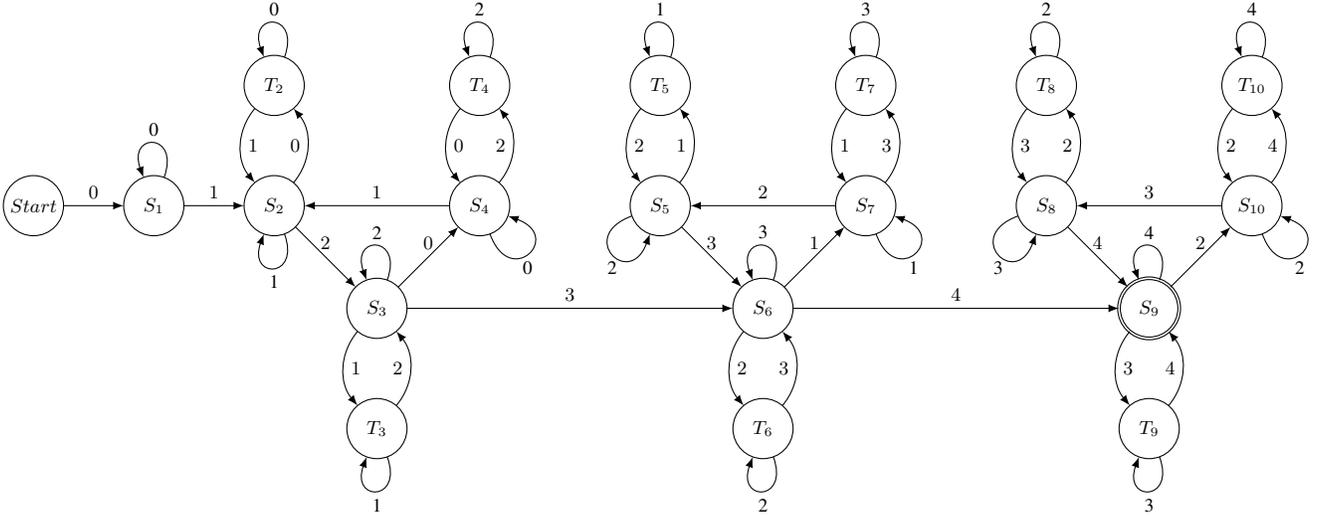


\begin{lemma}\label{lem:abc_IS_upperk}
For any alphabet $\Sigma_{q}$,
\begin{align}
&\max_{\bx\in\Sigma^3_{q}}\ \max_{\bx''\in D^{*,1}(\bx)} |R(\bx'')|= 13,\label{eq:irr3}\\
&\max_{\bx\in\Sigma^5_{q}}\ \max_{\bx''\in D^{*,1}(\bx)} |R(\bx'')|\le 17.\label{eq:irr5}
\end{align}
\end{lemma}

\begin{IEEEproof}
For the first statement, we may assume the symbols of $\bx$ are distinct, and in particular, we may assume without loss of generality that $\bx=012$. To see this, consider $\bx$ with repeated symbols, e.g., $\bx=010$. After a given sequence of $\le\!\!3$-TDs and a substitution, we will obtain $\bx''$. We then deduplicate all repeats to obtain $R(\bx'')$. For the same sequence of errors, since any deduplication that is possible when $\bx=012$ is also possible when $\bx=010$, the length of $R(\bx'')$ is not larger for $\bx=010$ than it is for $\bx=012$. 
 Hence, from this point on, we assume $\bx=012$.

\begin{table}[]
    \centering
    \caption{ Paths representing irreducible strings starting from and ending at specific states.}
    \label{Table:Irreducible_Paths_012}
    \begin{tabular}{p{.4 cm}|p{2.80 cm}|p{5.4 cm}}
    \hline
         state & Irreducible paths \newline from `Start' to state  & Irreducible paths \newline from state to $S_3$  \\
        \hline
        $S_1$ & $0$ & $012$, $1012$, $12$, $12012$, \\
        \hline
        $S_2$ & $01$, $01201$ &  $012$,$1012$, $12$, $12012$, $2$, $2012$, $212$, $212012$ \\
        \hline
        $S_3$ & $012$ &  $012$, $02012$, $12$, $12012$, $2$, $2012$, $212$, $212012$ \\
        \hline
        $S_4$ & $0120$ &  $012$, $02012$, $1012$, $12$, $12012$, $2012$  \\
        \hline
        $T_2$ & $010$, $012010$ &  $012$, $1012$,$12$, $12012$ \\
        \hline
        $T_3$ & $0121$ &  $12$, $12012$, $2$, $2012$, $212$, $212012$\\
        \hline
        $T_4$ & $01202$ &   $012$, $02012$, $2012$\\
        \hline

    \end{tabular}
\end{table}

As shown in~\cite{jain2017capacity}, $D^{*}(\bx)$ is a regular language whose words can be described as paths from `Start' to $S_3$ in the finite automaton given in Figure~\ref{fig:FA_IS}, where the word associated with each path is the sequence of the edge labels. Let $\bx'\in D^*(\bx)$ and $\bx''\in D^{0,1}(\bx')$. 
Assume $\bx'=\bu w \bz$ and $\bx''=\bu \hat w \bz$, where $\bu,\bz$ are strings and $w$ and $\hat w$ are distinct symbols. The string $\bu$ represents a path from `Start' to some state $U$ and the string $\bz$ represents a path from some state $Z$ to $S_3$ in the automaton, where there is an edge with label $w$ from $U$ to $Z$.

Since $R(\bx'') = R(R(\bu)\hat w R(\bz))$, we have $|R(\bx'')|\le |R(\bu)|+1+|R(\bz)|$. The maximum value for $|R(\bu)|$ is the length of some path from `Start' to $U$ such that the corresponding sequence does not have any repeats (henceforth, called an \emph{irreducible path}). All such paths/sequences are listed in the second column of Table~\ref{Table:Irreducible_Paths_012} for all choices of $U$. Similarly, the maximum value for $|R(\bz)|$ is the length of some irreducible path from $Z$ to $S_3$; all such possibilities are listed in the third column of  Table~\ref{Table:Irreducible_Paths_012}. An inspection of Table~\ref{Table:Irreducible_Paths_012} shows that choosing $U=T_2$ and $Z=S_2$ leads to the largest value of $|R(\bu)|+1+|R(\bz)|$, namely $6+1+6=13$. We note that the specific sequence achieving this length is $\bx''=0120103212012$, which can be obtained via the sequence
$\bx\to
012\,\underline{012}\,\underline{012}\to
012\,01\underline{01}2\,012\to012\,01012\underline{12}\,012\to\bx''$, where we have combined non-overlapping duplications into a single step.

Let us now prove the second statement.
Again we need only consider $\bx=01234$, for which $D^*(\bx)$ is the regular language whose automaton is shown in Figure~\ref{fig:fsm5}. 
In a similar manner to the proof of the previous part, we can show that the length of the longest irreducible path from `Start' to any state in the automaton is at most $8$ and the length of the longest irreducible path from any state to $S_9$ is also at most $8$. Hence, $|R(\bx'')|\le 8+1+8 = 17$, completing the proof. 
\end{IEEEproof}

We now consider changes to the roots of arbitrary strings when passed through a channel with arbitrarily many $\le\!\!3$-TDs and one substitution. {The next lemma is used in the main result of this section, Theorem~\ref{Them:error_IS_upperk}, which shows that even though a substituted symbol may be duplicated many times, the effect of a substitution on the root is bounded.}

\begin{lemma}\label{lem:decomposition}
Let $\bx$ be any string of length at least 5 and $\bx'\in D^*(\bx)$. 
For any decomposition of $\bx$ as
\[\bx=\ \br \ ab\ \bt \ de\ \bs,\]
for $a,b,d,e \in \Sigma_{q}$ and $\br, \bt,\bs \in \Sigma^{*}_{q}$, with $\bt$ nonempty, there is a decomposition of $\bx'$ as \[\bx'=\bu\  ab\ \bw \ de \ \bv\] such that $\bu, \bw,\bv \in \Sigma^{*}_{q}$,
 $\bu ab\in D^*(\br ab)$, $ab\bw de\in D^*(ab\bt de)$, and $de\bv\in D^*(de\bs)$. 
\end{lemma}
\begin{IEEEproof}
    If $\bx=\bx'$, the claim is true since we may choose $\bu=\br, \bw=\bt,\bv=\bs$. It suffices to consider the case in which $\bx'$ is obtained from $\bx$ via a single duplication. The case of more  duplications can be proved inductively.

    First suppose the length of the duplication transforming $\bx$ to $\bx'$ is 1. If this duplication occurs in $\br$, we choose $\bu$ to be the descendant of $\br$ and let $\bw=\bt$ and $\bv=\bs$, satisfying the claim. Duplication of a single symbol in $\bt$ or $\bs$ is handled similarly. If $a$ is duplicated, we let $\bu=\br a$, $\bw=\bt$, $\bv=\bs$. If $b$ is duplicated, we let $\bu=\br$, $\bw=b\bt$, $\bv=\bs$. The cases for $d$ and $e$ are similar.

    Second, consider a duplication of length 2 or 3. Such a duplication is fully contained in $\br ab$, $ab\bt de$, or $de \bs$. A duplication of length 2 or 3 applied to a string $\bz$ does not alter the first two and the last two symbols of $\bz$. So, for example, if the duplication occurs in $\br ab$, then we can choose $\bu$ such that $\bu ab\in D^1(\br ab)$ and let $\bw=\bt$ and $\bv=\bs$. The cases of duplications contained in the other strings are similar.
\end{IEEEproof}

\begin{theorem}\label{Them:error_IS_upperk}
Let $\cL$ be the smallest integer such that for any alphabet $\Sigma_{q}$, any $\bx\in\Sigma_{q}^*$, and any $\bx''\in D^{*,1}(\bx)$, we can obtain $R(\bx'')$ from $R(\bx)$ by deleting a substring of length at most $\cL$ and inserting a substring of length at most $\cL$ in the same position. Then $\cL\le 17$. 
\end{theorem}

\begin{IEEEproof}
We may assume $\bx$ is irreducible. If it is not, let $\bx_0=R(\bx)$ so that $\bx''\in D^{*,1}(\bx)\subseteq D^{*,1}(\bx_0)$. If the statement of the theorem holds for $\bx_0$, it also holds for $\bx$ since $R(\bx)=R(\bx_0)$.

We will find $\balpha,\bbeta,\bbeta',\bgamma\in\Sigma^*_{q}$ with $R(\bx)=\balpha\bbeta\bgamma$ and $R(\bx'')=\balpha\bbeta'\bgamma$ such that $|\bbeta'|\le 17$. Note that it suffices to prove $|\bbeta'|\le 17$ for all irreducible $\bx$. To see this, note that $\balpha\bbeta'\bgamma$ is obtained from $\balpha\bbeta\bgamma$ by applying, in order, duplications, a single substitution, more duplications, and finally removing all repeats (performing all possible deduplications). Since duplications that occur after the substitution do not make any difference, we may instead assume that the process is as follows: duplications, substitution, deduplications. Since this process is reversible, general statements that hold for {$\bbeta'$} also hold for {$\bbeta$}.


Let $\bx'\in D^*(\bx)$ be obtained from $\bx$ through duplications and $\bx''$ be obtained from $\bx'$ through a substitution. We assume that $\bx=\br abcde\bs$, where $\br,\bs\in\Sigma^*_{q}$ and  $a,b,c,d,e\in\Sigma_{q}$, such that the substituted symbol in $\bx'$ is a copy of $c$. Note that if $|\bx|<5$ or if a copy of one of its first two symbols or its last two symbols are substituted, then we can no longer write $\bx$ as described. To avoid considering these cases separately, we may append two dummy symbols to the beginning of $\bx$ and two dummy symbols to the end of $\bx$, where the four dummy symbols are distinct and do not belong to $\Sigma_{q}$, and prove the result for this new string. Since these dummy symbols do not participate in any duplication, substitution, or deduplication events, the proof is also valid for the original $\bx$.

With the above assumption and based on Lemma~\ref{lem:decomposition},
we can write
\begin{equation}\label{eq:decomposition}
    \begin{split}
    \bx&=\ \br \ ab\ c \ de\ \bs\\
    \bx'&=\bu\  ab\ \bw \ de \ \bv \in D^*(\bx),\\
    \bx'' &= \bu \ ab\ \bz \ de \ \bv \in D^{0,1}(\bx'),
    \end{split}
\end{equation}
where $\bu ab\in D^*(\br ab)$, $ab\bw de\in D^*(abc de)$,  $de\bv\in D^*(de\bs)$, and $\bz$ is obtained from $\bw$ by substituting an occurrence of $c$.
From~\eqref{eq:decomposition}, $R(\bx'')=R( \br R(ab \bz de ) \bs )$, where $R(ab \bz de )$ starts with $ab$ and ends with $de$ (which may fully or partially overlap). The outer $R$ in $R( \br R(ab \bz de ) \bs )$ may remove some symbols at the end of $ \br $, beginning and end of $R(ab \bz de )$, and the beginning of $ \bs $, leading to $\balpha\bbeta'\bgamma$, where $\balpha$ is a prefix of $ \br $, $\bbeta'$ is a substring of $R(ab \bz de )$, and $\bgamma$ is a suffix of $ \bs $. Hence, $|\bbeta'|\le|R(ab \bz de )|$. But $ab\bz de\in D^{*,1}(abcde)$ and thus by Lemma~\ref{lem:abc_IS_upperk}, $|R(ab\bz de)|\le17$, completing the proof. 
\end{IEEEproof}

We provide an example for Theorem~\ref{Them:error_IS_upperk}, where the root of a sequence is altered by several duplications and one substitution.

\begin{example}\label{Exp:shrt_TD_substitution_examples}
Fix $\Sigma_{4}=\{0,1,2,3\}$ as the alphabet. In the following examples, $\bx$ is an irreducible string, $\bx'\in D^*(\bx)$, and $\bx''\in D^{0,1}(\bx')$. We compare $R(\bx)=\bx$ with $R(\bx'')$.

\begin{itemize}
    \item Let $\bx=012302$,  $\bx'=01\underline{1}2 \underline{012}\underline{012} 302\underline{02}$, and $\bx''=01\underline{1}2\underline{01\bcomment{\bm{3}}}\underline{012}302\underline{02}$, where the underlined symbols result from duplication  and the bold symbol from  substitution. Then $R(\bx'')=012013012302$ and the change from $R(\bx)$ to $R(\bx'')$ can be viewed as
    \begin{equation*}
        R(\bx)=\underbrace{012}_{\balpha}\underbrace{302}_{\bgamma}\to R(\bx'')=\underbrace{012}_{\balpha}\underbrace{013012}_{\bbeta'}\underbrace{302}_{\bgamma},
    \end{equation*}
    with $\bbeta = \Lambda$.

        \item Let $\bx=13203103$,  $\bx'=13\underline{13}2\underline{132}0310\underline{310}3$, and $\bx''=13\underline{13}2\underline{13\bcomment{\bm{1}} }0310\underline{310}3$. Then $R(\bx'')=13213103$ and the change from $R(\bx)$ to $R(\bx'')$ can be viewed as
    \begin{equation*}
        R(\bx)=\underbrace{132}_{\balpha}\underbrace{0}_{\bbeta}\underbrace{3103}_{\bgamma}\to R(\bx'')=\underbrace{132}_{\balpha}\underbrace{1}_{\bbeta'}\underbrace{3103}_{\bgamma}.
    \end{equation*}

     \item Let $\bx=012010321201230$,  $\bx'=012\underline{012}01032120\underline{20}12\underline{012}30$, and $\bx''=012\underline{012}010\bcomment{\bm{1}}2120\underline{20}12\underline{012}30$. Then $R(\bx'')=01230$ and the change from $R(\bx)$ to $R(\bx'')$ can be viewed as
    \begin{equation*}
        R(\bx)=\underbrace{012}_{\balpha}\underbrace{0103212012}_{\bbeta}\underbrace{30}_{\bgamma}\to R(\bx'')=\underbrace{012}_{\balpha}\underbrace{30}_{\bgamma},
    \end{equation*}
    with  $\bbeta' = \Lambda$.

\end{itemize}
\end{example}


\section{Error-correcting codes} \label{sec:ECC}
Having studied how duplication roots are affected by tandem duplication and substitution errors,  we now construct codes that can correct such errors. We will also determine the rate of these codes and compare it with the rate of codes that only correct duplications, which provides an upper bound. 



\subsection{Code constructions}
As noted in the previous section, the effect of a substitution error on the root of the stored codeword is local in the sense that a substring of bounded length may be deleted and another substring of bounded length may be inserted in its position. A natural approach to correcting such errors is to divide the codewords into blocks such that this alteration can affect a limited number of blocks. In particular, we divide the string into \emph{message blocks} that are separated by \emph{marker blocks} known to the decoder. We start with an auxiliary construction.

\begin{construction}\label{cons:blocks_separated}
Let $l,m,N$ be positive integers with $m>l$ and $\bsigma\in \ir(l)$. The code $\cC_{\bsigma}$ (where dependence on $N,m$ is implicit) consists of strings $\bx$ obtained by alternating between \emph{message blocks} of length $m$ and copies of the \emph{marker} sequence $\bsigma$, i.e.,
\[\bx = B_1\bsigma B_2 \bsigma \dotsm \bsigma B_N,\]
such that $\bx\in \ir(N(m+l)-l)$, $B_i\in  \ir(m)\subseteq \Sigma^m_{q}, i\in[N]$, and there are exactly two occurrences of $\bsigma$ in ${\bsigma}B_i\bsigma$, for all $i\in[N]$. (Thus, there are precisely $N-1$ occurrences of $\bsigma$ in $\bx$.)
\end{construction}


We remark that for our purposes, we can relax the condition on $\bsigma B_i\bsigma$ for $i=1,N$. Specifically, it suffices to have exactly one occurrence of ${\bsigma}$ in $B_1{\bsigma}$ and one occurrence of $\bsigma$ in $\bsigma B_N$. For simplicity however, we do not use these relaxed conditions.

With this construction in hand, in the next theorem, we show that the effect of one substitution and many tandem duplications is limited to a small number of blocks.

\begin{theorem}\label{Theorem:error_blocks_update}
Let $\cC_{\bsigma}$ be the code defined in Construction~\ref{cons:blocks_separated}. If $m>\cL$, then there exists a decoder $\cD_{\bsigma}$ that, for any $\bx\in\cC_{\bsigma}$ and $\by\in R(D^{*,\le1}(\bx))$, outputs $\bz=\cD_{\bsigma}(\by)$ such that, relative to $\bx$, either two of the blocks $B_i$ are substituted in $\bz$ or four of them are erased.
\end{theorem}
\begin{IEEEproof}
Let $\bx=\balpha\bbeta\bgamma$ and $\by = \balpha\bbeta'\bgamma$, where by Theorem~\ref{Them:error_IS_upperk}, $|\bbeta|,|\bbeta'|\le \cL$. The decoder considers two cases depending on whether the marker sequences $\bsigma$ are in the same positions in $\by$ as in the codewords in $\cC_{\bsigma}$. If this is the case, then $|\bbeta|=|\bbeta'|\le \cL$. Since $\cL<m=|B_i|$, at most two (adjacent) blocks $B_i$ are affected by substituting $\bbeta$ by $\bbeta'$ and thus $\bz=\by$ differs from $\bx$ in at most two blocks.

On the other hand, if the markers are in different positions in $\by$ compared to the codewords in $\cC_{\bsigma}$, the decoder uses the location of the markers to identify the position of the blocks that may be affected and erases them, as described below. To avoid a separate treatment for blocks $B_1$ and $B_N$, the decoder appends ${\bsigma}$ to the beginning and end of $\by$ and assumes that the codewords are of the form ${\bsigma}B_1{\bsigma}\dotsm{\bsigma}B_N{\bsigma}$.

Define a block in $\by$ as a maximal substring that does not overlap with any ${\bsigma}$. By the assumption of this case, there is at least one block $B$ in $\by$ whose length differs from $m$. Hence, $\by$ has a substring $\bu$ of length $m+2l$ that starts with ${\bsigma}$ and contains part or all of $B$ but does not end with ${\bsigma}$.

Since $\bu$ is not a substring of any codeword, it must overlap with $\bbeta'$ (if $\bbeta'$ is the empty string, then $\bu$ surrounds the location from which $\bbeta$ was deleted and we may still consider that $\bu$ and $\bbeta'$ overlap). Let ${\delta}=|\bx|-|\by|=|{\bbeta}|-|{\bbeta'}|$ and ${\delta}^+=\max(0,{\delta})$. Note that $|\bbeta'|=|\bbeta|-\delta \le \cL - \delta$. Since $|{\bbeta}'|\le \min(\cL,\cL-{\delta})=\cL-{\delta}^+$, removing $\bu$ along with the $\cL-{\delta}^+-1$ elements on each of its sides, with a total length of $m+2l+2\cL-2{\delta}^+-2$, will remove $\bbeta'$ from $\by$. This results in a sequence that relative to $\bx$ suffers a deletion of length at most $m+2l+2\cL-2{\delta}^+-2+|{\bbeta}|-|{\bbeta}'|<3m+2l$ from a known position. The deletion affects at most 4 blocks and since its location is known, the decoder can mark these blocks as erased.
\end{IEEEproof}

In Construction~\ref{cons:blocks_separated}, the constraint that $\bx$ must be irreducible creates interdependence between the message blocks, making the code more complex. The following lemma allows us to treat each message block independently provided that ${\bsigma}$ is sufficiently long.

\begin{lemma}\label{Them:separation_block_length_updated}
Let $\bx$ be as defined in Construction~\ref{cons:blocks_separated} and assume $l\ge5$. The condition $\bx\in \ir(N(m+l)-l)$ is satisfied if
\begin{equation}\label{eq:indep}
    {\bsigma}B_i{\bsigma}\in \ir(m+{2}l), \quad\text{for all } i\in[N].
\end{equation}
\end{lemma}
\begin{IEEEproof}
Suppose that $\bx$ has a repeat $\ba\ba$, with $|\ba|\le 3$. Since $|\ba\ba|\le6$ and $|\bsigma|\ge5$, there is no $i$ such that the repeat lies in $B_i{\bsigma}B_{i+1}$ and overlaps both $B_i$ and $B_{i+1}$. So it must be fully contained in $B_1{\bsigma}$, ${\bsigma}B_N$, or ${\bsigma}B_i{\bsigma}$ for some $2\le i\le N-1$,
contradicting assumption~\eqref{eq:indep}.
\end{IEEEproof}

We now present a code based on Construction~\ref{cons:blocks_separated} and prove that it can correct any number of tandem duplications and one substitution error.
\begin{construction}\label{cons:error_RS_codes_updated}
Let $l,m$ be positive integers with $m>l\ge5$, and ${\bsigma}\in \ir(l)$. Furthermore, let $\cB_{\bsigma}^m$ denote the set of sequences $B$ such that ${\bsigma}B{\bsigma}\in \ir(m+2l)$ has exactly two occurrences of ${\bsigma}$, and $M=M^{(m)}_{\bsigma}=|\cB_{\bsigma}^m|$. Finally, let $t$ be a positive integer such that $2^t\le M$ and $\zeta:\bbF_{2^t} \to \cB_{\bsigma}^m$ be a one-to-one mapping. We define $\cC_{MDS}$ as
\begin{equation*}
    \cC_{MDS} = \{\zeta(c_1){\bsigma}\zeta(c_2){\bsigma}\dotsm{\bsigma}\zeta(c_N)\!:\!\bc\in \mds(N,N-4,5)\},
\end{equation*}
where $\mds(N,N-4, 5 )$ denotes an MDS code over $\bbF_{2^t}$ of length $N=2^t-1$, dimension $N-4$, and Hamming distance $d_H=5$.
\end{construction}

\begin{theorem}\label{Them:ECC_update}
If {$m>\cL$,} then the error-correcting code $\cC_{\mds}$ in Construction~\ref{cons:error_RS_codes_updated} can correct any number of $\le\!\!3$-TD and at most one substitution errors.
\end{theorem}

\begin{IEEEproof}
Let the stored codeword be $\bx=B_1{\bsigma}\dotsm{\bsigma}B_N\in\cC_{MDS}$, where $B_i=\zeta(c_i)$ for $i\in[N]$ and $\bc\in C$, with $C$ denoting an $\mds(N,N-4,5)$ code. Suppose the retrieved word is $\by$. By Lemma~\ref{Them:separation_block_length_updated}, $\cC_{MDS}\subseteq \cC_{\bsigma}$. By Theorem~\ref{Theorem:error_blocks_update}, $\cD_{\bsigma}(\by)$ suffers either at most two substitutions or at most four erasures of blocks. Suppose block $B_i$ is substituted by another string $\bv$ of length $m$. If $\zeta^{-1}(\bv)$ exists, this translates into a substitution of $c_i$. If not, we define $\zeta^{-1}(B_i)$ as an arbitrary element of $\bbF_{2^t}$, again leading to a possible substitution of $c_i$ with another symbol. To decode, we can use the MDS decoder on $\zeta^{-1}(\cD_{\bsigma}(\by))$, which relative to $\bc$ suffers either $\le2$ substitutions or $\le4$ erasures. Given that the minimum Hamming distance of the MDS code is 5, the decoder can successfully recover $\bc$.
\end{IEEEproof}



\subsection{Construction of message blocks}
In this subsection, we study the set $\cB_{\bsigma}^m$ of valid message blocks of length $m$ with ${\bsigma}$ as the marker. Since in Construction~\ref{cons:error_RS_codes_updated}, the markers ${\bsigma}$ do not contribute to the size of the code, to maximize the code rate, we set $l=|\bsigma|=5$, i.e., $\bsigma \in \ir(5)$.

For a given ${\bsigma}$, we need to find the set $\cB_{\bsigma}^m$. The first step in this direction is finding all irreducible sequences of length $m+2l=m+10$. We will then identify those that start and end with ${\bsigma}$ but contain no other ${\bsigma}$s. 

As shown in~\cite{jain2017duplication}, the set of $\leq\!\!3$-irreducible strings over an alphabet of size $q$ is a regular language whose graph $G_q=(V_{q},\xi_{q})$ is a subset of the De Bruijn graph. The vertex set $V_q$ consists of 5-tuples $a_1a_2a_{3}a_{4}a_5$ that do not have any repeats (of length at most 2). There is an edge from $a_{1}a_{2}a_{3}a_{4}a_{5} \to a_2a_3a_4a_5a_6$ if $a_{1}a_{2}a_{3}a_{4}a_{5}a_{6}$ belongs to $\ir(6)$. The label for this edge is $a_6$. The label for a path is the 5-tuple representing its starting vertex concatenated with the labels of the subsequent edges. In this way, the label of a path in this graph is an irreducible sequence 
and each irreducible sequence is the label of a unique path in the graph
. The graph $G_q$, when $q=3$, can be found in \cite[Fig.~1]{jain2017duplication}.

The following theorem characterizes the set $\cB_{\bsigma}^m$ and will be used in the next section to find the size of the code.

\begin{theorem}\label{Them:block_paths}
Over an alphabet of size $q$ and for ${\bsigma}\in \ir(5)$, there is a one-to-one correspondence between 
{$B\in \cB_{\bsigma}^m$} and paths of length $m+5$ in $G_q$ that start and end in ${\bsigma}$ but do not visit ${\bsigma}$ otherwise. Specifically, each sequence $B\in\cB_{\bsigma}^m$ corresponds to the path with the label ${\bsigma}B{\bsigma}$.
\end{theorem}

\begin{IEEEproof}
Consider a path $p=\bv_1\bv_2\dotsm \bv_{k+1}$ where $\bv_i$ are vertices of $G_q$ and $k$ is the length of the path. Denote the label of this path by $\bs=s_1s_2\dotsm s_{k+5}$. It can be shown by induction on $k$ that $\bv_i = s_is_{i+1}s_{i+2}s_{i+3} s_{i+4}$. Hence, the label of a path of length $m+5$ that starts and ends in $\bsigma$ but does not visit $\bsigma$ otherwise is an irreducible sequence with exactly two occurrences of $\bsigma$ and is of the form $\bsigma B\bsigma$ where $B\in\cB_{\bsigma}^m$. Conversely, suppose $B\in\cB_{\bsigma}^m$. Then $\bsigma B\bsigma$ is an irreducible string of length $m+10$ and thus the label of a unique path of length $m+5$ in $G_q$. This path starts and ends in $\bsigma$. But it does not visit $\bsigma$ in its interior since that would imply there are more than two occurrences of $\bsigma$ in $\bsigma B\bsigma$.
\end{IEEEproof}
\subsection{Code rate}
We now turn to find the rate of the code introduced in this section. For a code $C$ of length $n$ and size $|C|$, the rate is defined as $R(C)=\frac1n\log|C|$. For the code of Construction~\ref{cons:error_RS_codes_updated}, 
\begin{equation}\label{eq:code_rate_MDS}
\begin{split}
       R(\cC_{\mds})&=\frac{N-4}{Nm+(N-1)l}\log(N+1),
\end{split}
\end{equation}
where $N$ depends on the choice of ${\bsigma}\in \ir(5)$. More specifically, $N\le 2^{\lfloor\log M_{\bsigma}^{(m)}\rfloor}-1$. Choosing the largest permissible value for $N$ implies that $N\ge({M_{\bsigma}^{(m)}-1})/{2}$ and
\begin{equation}\label{eq:code_rate_MDS_M}
\begin{split}
       R(\cC_{\mds})&\ge\frac{1-4/N}{m+l}\log(N+1)\\&\ge\frac{1}{m+l}\left(1-\frac{8}{M_{\bsigma}^{(m)}-1}\right)(\log M_{\bsigma}^{(m)}-1).
\end{split}
\end{equation}
If we let $m$ and $M_{\bsigma}^{(m)}$ grow large, the rate becomes
\begin{equation}\label{eq:assymp_rate}
    R(\cC_{\mds})=\frac1m{\log M_{\bsigma}^{(m)}}(1+o(1)).
\end{equation}


For a given alphabet $\Sigma_{q}$, let $A$ denote the adjacency matrix of $G_q$, where the rows and columns of $A$ are indexed by $\bv\in V_q\subseteq\Sigma_q^5$. Furthermore, let $A_{(\bv)}$ be obtained by deleting the row and column corresponding to $\bv$ from $A$ and $c_{(\bv)}$ (resp.\ $r_{(\bv)}^T$) be the column (row) of $A$ corresponding to $\bv$ with the element corresponding to $\bv$ removed. Recall that $M^{(m)}_{\bsigma}=|\cB_{\bsigma}^m|$. From Theorem~\ref{Them:block_paths}, we have
\begin{equation}\label{eq:M}
    M^{(m)}_{\bsigma}=r_{(\bsigma)}^T {\left(A_{(\bsigma)}\right)^{m+l-2}}c_{(\bsigma)}
\end{equation}
where $(\cdot)^T$ denotes matrix transpose.
As $m\to\infty$, if $A_{(\bsigma)}$ is primitive~\cite{lind1995introduction}, 
we have
\begin{equation}\label{eq:M_large}
    \frac1{m+l}\log M^{(m)}_{\bsigma}\to \log(\lambda_{\bsigma}),
\end{equation}
where $\lambda_{\bsigma}$ is the largest eigenvalue of $A_{({\bsigma})}$.
Maximizing over ${\bsigma}\in V_q$ yields the largest value for $M_{\bsigma}^{(m)}$ in~\eqref{eq:M} and~\eqref{eq:M_large}, and thus the highest code rate. This is possible to do computationally for small values of $q$ and, in particular, for $q=4$, which corresponds to data storage in DNA.
In this case, $A_{({\bsigma})}$ is primitive for all choices of ${\bsigma}\in \ir(5)$ and the largest eigenvalue is obtained for ${\bsigma}=01201$ (and strings obtained from $01201$ by relabeling the alphabet symbols). For this ${\bsigma}$, we find $\lambda_{\bsigma}=2.6534$, leading to an asymptotic code rate of $1.4078$ bits/symbol.

It was shown in~\cite{jain2017duplication} that the set of irreducible strings of length $n$ is a code correcting any number of $\le\!\!3$-TDs. In~\cite{kovavcevic2019codes}, it was shown that the rate of this code, $\frac1n\log|\ir(n)|$, is asymptotically optimal. It is easy to see that $\frac1n\log|\ir(n)|\le\log(q-1)$ as no symbol can be repeated. For the case of $q=4$, we have $\frac1n\log|\ir(n)|=\log 2.6590=1.4109$ bits/symbol. Therefore, the cost of protection against a single substitution in our construction is only $0.003$ bits/symbol. It should be noted, however, that here we have assumed $m$ is large, thus ignoring the overhead from the MDS code and marker strings.

In addition to the computational rate obtained above for the important case of $q=4$, we will provide analytical bounds on the code rate. An important quantity affecting the rate of the code is the number of outgoing edges from each vertex in $G_q$ that do not lead to $\bsigma$. The asymptotic rate of the code is bounded from below by the number of such edges. The next lemma, which establishes the number of outgoing edges for each vertex, will be useful in identifying an appropriate choice of ${\bsigma}$, and the following theorem provides a lower bound for $M_{\bsigma}^{(m)}$ for such a choice.
\begin{lemma}\label{lem:output_edges}
For $q>2$, a vertex $\bv=a_1a_2a_3a_4a_5$ in $G_q$ has $q-2$ outgoing edges if $a_3=a_5$ or $a_1a_2=a_4a_5$. Otherwise, it has $q-1$ outgoing edges.
\end{lemma}

\begin{IEEEproof}
    Consider $\bv=a_1a_2a_3a_4a_5\in\ir(5)$, and $\bw=a_2a_3a_4a_5a_6\in\ir(5)$. There is an edge from $\bv$ to $\bw$ if $a_1a_2a_3a_4a_5a_6\in\ir(6)$. The number of outgoing edges from $\bv$ equals the number of possible values for $a_6$ such that this condition is satisfied. Clearly, $a_6\neq a_5$. Furthermore, if $a_3 = a_5$, then $a_6\neq a_4$ and if $a_1a_2=a_4a_5$, then $a_6\neq a_3$.

    However, $a_3 = a_5$ and $a_1a_2=a_4a_5$ cannot simultaneously hold, since that would imply $a_2=a_3$, contradicting $\bv\in\ir(5)$. Hence, if either $a_3=a_5$ or $a_1a_2=a_4a_5$ holds, then there are $q-2$ outgoing edges and if neither holds, there are $q-1$ outgoing edges.
\end{IEEEproof}


Since $\bsigma$ must also be excluded, it may seem that the number of outgoing edges may be as low as $q-3$. But we show in the next theorem that with an appropriate choice of $\bsigma$, we can have $q-2$ as the lower bound.

\begin{theorem}\label{them:lower_bound_M}
Over an alphabet of size $q>2$, there exists ${\bsigma}\in\ir(5)$ such that
\(
M_{\bsigma}^{(m)}\ge (q-2)^{m-c_q},
\)
where $c_q$ is a constant independent from $m$.
\end{theorem}

\begin{IEEEproof}
Recall that $M_{\bsigma}^{(m)}$ is the number of paths of length $m+5$ in $G_q$ that start and end in ${\bsigma}$ but do not visit ${\bsigma}$ otherwise. Since the path must return to ${\bsigma}$, we will show below that for an appropriate choice of ${\bsigma}$, there is a path in $G_q$ from any vertex to ${\bsigma}$, and define $c_q$ such that the length of this path is at most $c_q+5$. Hence $M_{\bsigma}^{(m)}$ is at least the number of paths of length $m-c_q$ from ${\bsigma}$ to another vertex that do not pass through ${\bsigma}$.

As shown in Lemma~\ref{lem:output_edges}, each vertex in $G_q$ has at least $q-2$ outgoing edges. We select ${\bsigma}$ such that this still holds even if edges leading to ${\bsigma}$ are excluded. We do so by ensuring that each vertex $\bv$ with an outgoing edge to ${\bsigma}$ has $q-1$ outgoing edges. Let $\bv=a_1a_2a_3a_4a_5$ and ${\bsigma}=a_2a_3a_4a_5a_6$. Based on Lemma~\ref{lem:output_edges}, if $a_2\neq a_5$ and $a_3\neq a_5$, then $\bv$ has $q-1$ outgoing edges. In particular, we can choose ${\bsigma}=01020$ since $q\ge3$. With this choice, $M_{\bsigma}^{(m)}\ge(q-2)^{m-c_q}$.

To complete the proof, we need to show that there is a path in $G_q$ from any vertex to ${\bsigma}=01020$. For $q=3,4,5$, we have checked this claim computationally by explicitly forming $G_q$. Let us then suppose $q\ge 6$, where the alphabet $\Sigma_q$ contains $\{3,4,5\}$. Let $\bv=a_1\dotsm a_5$ be some vertex in $G_q$. There is an edge from $\bv$ to $a_2\dotsm a_6$ for some $a_6\in\{3,4,5\}$ since, from Lemma~\ref{lem:output_edges}, at most two elements of $\Sigma_q$ are not permissible. Continuing in similar fashion, in 5 steps, we can go from $\bv$ to some vertex $\bw=b_1\dotsm b_5$ whose elements $b_i$ belong to $\{3,4,5\}$. We can then reach ${\bsigma}$ in 5 additional steps via the path $\bw\to b_2\dotsm b_4b_50\to b_3b_4b_501\to \dotsm\to {\bsigma}$, proving the claim. In particular, for $q\ge6$, we have $c_q\le5$.
\end{IEEEproof}

We can now find a lower bound on the asymptotic rate, based on \eqref{eq:assymp_rate} and the proceeding theorem:
\begin{corollary}
For $q>2$, as $m\to\infty$,
\(
R(\cC_{\mds})\ge \log(q-2)(1+o(1)).
\)
\end{corollary}
We note that this gives the lower bound of 1 bit/symbol for $q=4$, which we can compare to the upper bound of $\log (q-1)=1.585$ for codes correcting only duplications and to the rate obtained computationally following~\eqref{eq:M_large}, which was 1.4078 bits/symbol.


\section{Conclusion}\label{sec:conclusion}

This paper considered constructing error-correcting codes for channels with many short duplications and one unrestricted substitution error. Because  the channel allows  an arbitrary number of duplications, a single substitution may affect an unbounded segment of the output,  as the substituted symbol may appear many times in different positions. However, with an appropriate construction of message blocks and processing of the output strings, the substitution error leads to the erasure of at most  $4$ message blocks or substitution of at most 2. Therefore, a maximum distance separable (MDS) code with minimum Hamming distance $5$ over message blocks can correct these errors. However, there is an additional requirement. Namely, the codewords must be irreducible. Separating the message blocks with a marker sequence $\bsigma$ of length at least 5 allows us to ensure that the codewords are repeat-free by guaranteeing that each message block is irreducible. The rate of the code is determined by the number of such blocks, which in turn depends on the marker sequence $\bsigma$. We showed that permitted message blocks are paths in a modified De Bruijn graph and that choosing $\bsigma$ appropriately allows each vertex to have at least $q-2$ outgoing edges, thus guaranteeing an asymptotic rate of at least $\log(q-2)$. 
When $q=4$, the case corresponding to DNA storage, a computational bound for the code rate shows that the asymptotic rate is only $0.003$ bits/symbol smaller than that of the code that corrects short duplications but no substitutions.

It remains an open problem to efficiently correct more substitution errors. Another, possibly more challenging, problem is correcting substitutions and duplications of length bounded by an arbitrary constant $k$. If $k$ is larger than $3$, the duplication root is no longer unique \cite{jain2017duplication}, which complicates the code design. Furthermore, a key feature of duplications of length at most $3$ is that such duplications  lead to regular languages. We used this fact to characterize the effect of the channel on the roots of sequences. However, if $k\geq 4$, then the language is not regular \cite{leupold2003formal}, leading to challenges in characterizing the channel.

\bibliographystyle{IEEEtran}
\bibliography{references}

\end{document}